\documentclass[twocolumn,showpacs,preprintnumbers,amsmath,amssymb,superscript address, revsymb,prb]{revtex4}

\usepackage[dvips]{graphicx}
\usepackage{mathptmx}
\usepackage{dcolumn}

\newcommand{\etal}{\emph{et al.}}
\newcolumntype{.}{D{.}{\cdot}{3.10}}

\begin{document}

\title{Energetics of hydrogen/lithium complexes in silicon analyzed
  using the Maxwell construction}

\author{Andrew J. Morris\footnote{Email: andrew.morris@ucl.ac.uk.}}
\affiliation{Department of Physics and Astronomy, University College
  London, Gower Street, London WC1E 6BT, United Kingdom}

\author{C. P. Grey} \affiliation{Department of Chemistry, University
  of Cambridge, Lensfield Road, Cambridge CB2 1EW, United Kingdom}

\author{R. J. Needs} \affiliation{Theory of Condensed Matter Group,
  Cavendish Laboratory, University of Cambridge, J.~J. Thomson Avenue,
  Cambridge CB3 0HE, United Kingdom}

\author{Chris J. Pickard} \affiliation{Department of Physics and
  Astronomy, University College London, Gower Street, London WC1E 6BT,
  United Kingdom}

\date{\today{}}

\begin{abstract}
  We have studied hydrogen/lithium complexes in crystalline silicon
  using density-functional-theory methods and the \textit{ab initio}
  random structure searching (AIRSS) method for predicting structures.
  A method based on the Maxwell construction and convex hull diagrams
  is introduced which gives a graphical representation of the relative
  stabilities of point defects in a crystal and enables visualization
  of the changes in stability when the chemical potentials are
  altered.  We have used this approach to study lithium and hydrogen
  impurities in silicon, which models aspects of the anode material in
  the recently-suggested lithium-ion batteries.  We show that hydrogen
  may play a role in these anodes, finding that hydrogen atoms bind to
  three-atom lithium clusters in silicon, forming stable \{H,3Li\} and
  \{2H,3Li\} complexes, while the \{H,2Li\} complex is almost stable.
\end{abstract}

\maketitle

\section{Introduction}

The properties of a material may be substantially affected by the
presence of impurity atoms.\cite{stoneham:OUP:1975} One of the
fundamental properties is the relative energies of different
configurations of the impurities within the material.  Several
different species of atom may be present, and the impurity atoms may
have strong interactions with each other and with the host material.
In this paper we study the interactions between lithium and hydrogen
impurities in crystalline silicon, which is a system of interest for
lithium-ion battery technologies.  Lithium-ion batteries are widely
used in portable electronics and are now being employed in the next
generation of hybrid and all-electric vehicles. \cite{LIB_background}
Recently there has been interest in using silicon anodes, which have a
very high volumetric and gravimetric
capacity. \cite{lai:JES:1976,wen:JSSC:1981,weydanz:JPS:1999}
Crystalline silicon is generally used in the battery and the first
charge-discharge cycle involves the conversion of crystalline silicon
to an amorphous phase.

Point defects may form a variety of complexes containing several atoms
of more than one atomic species.  An interesting feature of our
approach is the use of a graphical representation of the relative
stabilities of the defects based on the Maxwell
construction\cite{gibbs_1873,maxwell_1874} and the corresponding
convex hull.  This is a standard procedure used in materials science
(and elsewhere) for studying phase separation, which we have applied
to defects in crystals.  The term ``convex hull'' refers to the fact
that the second law of thermodynamics requires that the energy per
particle (or free energy as appropriate) must be a convex function of
the relative concentrations of the particles.  Appropriate chemical
potentials for the different atomic species can normally be estimated
for a particular set of external conditions.  We are also interested
in the changes in the relative stabilities of defects when the
chemical potentials of the atomic species are altered.  Charged
defects are also of interest in general, and these can be included by
introducing a chemical potential for electrons, although we will not
show results for this case here.

This article is divided into the following sections, in
Sect.\,\ref{Sect:hull}, the modified Maxwell construction is
introduced to visualize the energetics of the phase separation of
defect complexes, and the example of N/O complexes in silicon is used
to validate the method.  In Sect.\,\ref{Sect:example} the method is
applied to the problem of H/Li defect complexes in silicon.  Finally
Sect.\,\ref{Sect:conclusions} concludes the work.

\section{Convex Hull for defects}
\label{Sect:hull}

In this article we denote a defect complex by listing its constituent
atoms between braces, for example, the \{2H,3Li\} defect complex
contains two hydrogen and three lithium atoms.  For simplicity we
begin by considering a single impurity species $\alpha$ which forms a
single impurity complex in a large sample of perfect bulk crystal.
The formation energy of a complex \{$n_{\alpha} \alpha$\} containing
$n_{\alpha}$ impurity atoms is given by
\begin{equation}
  E_f = E_{\mathrm{D}} - n_{\alpha}\mu_{\alpha} - E_{\mathrm{H}},
\label{eqn:form_eng}
\end{equation}
where $E_{\mathrm{D}}$ is the energy of the system including the
defect complex, $\mu_{\alpha}$ is the chemical potential of the
impurity species, and $E_{\mathrm{H}}$ is the energy of the host
crystal without defects.  The energy of the host crystal can be
written as
\begin{equation}
  E_{\mathrm{H}} = \displaystyle \sum_{\beta} n_{\beta}\mu_{\beta},
\label{eqn:form_EB}
\end{equation}
where there are $n_{\beta}$ atoms of the host crystal atoms of each species
$\beta$ which have chemical potentials $\mu_{\beta}$.  The complex with the
lowest value of $E_f/n_{\alpha}$ is the most stable and ones with
higher energies are metastable.

Eq.\,(\ref{eqn:form_eng}) can readily be extended to a complex
containing a larger number of impurity species, and the formation
energy per impurity atom is then 
\begin{equation}
  E_{\rm pa} = \frac{E_f}{\sum_{\alpha} n_{\alpha}} = \frac{E_{\mathrm{D}} - 
    \sum_{{\alpha}} n_{{\alpha}}\mu_{{\alpha}} - E_{\mathrm{H}}} 
  {\sum_{{\alpha}} n_{{\alpha}}}.
\end{equation}

The convex hull diagram is constructed by plotting the formation
energy per impurity atom of each complex against the fractional
concentration $C_{i}$ of impurity species $i$, where
\begin{equation}
  C_{i}=\frac{n_{i}}{\sum_{\alpha} n_{\alpha}}.
\end{equation}
If the complexes contain only two atomic species we can simply plot
$E_{\rm pa}(C_{i})$ on a graph as shown, for example, in Fig.\
\ref{Fig:NO}.  The solid red and blue lines in Fig.\ \ref{Fig:NO} show
convex hulls which consist of straight-line segments known as ``tie
lines''.  A higher-dimensional hull is required if there are more than
two impurity species.  The case of three atomic species can be
represented on a ``triangular plot'' as shown, for example, in Ref.\
\onlinecite{ong:cm:2008}.  Alternatively one can consider
two-dimensional slices of the multi-dimensional hull.  Electrons can
also be included as a species, so that defect complexes in different
charge states can be described.

When more than one impurity species is present it is possible for
``phase separation'' or ``disproportionation'' of defect complexes to
occur.  For example, the lowest energy state when equal numbers of
impurity species A and B are present might consist of equal numbers of
the defect complexes \{A,2B\} and \{A\}, rather than the defect
complex \{2A,2B\}.  We will assume that the defect complexes do not
interact with one another (except by a ``chemical reaction'', forming
new complexes which can be included within the theory) and they are
expected to be distributed randomly over the host crystal rather than
forming spatially distinct phases as in the standard application to
``phase separation''.
We may include within $E_{\rm pa}$ the contributions to the free energy from
the vibrational entropy and the configurational entropy due to the
different degenerate orientations of a complex at a lattice site.
However, it is not possible to include the configurational entropy due
to arranging defects at different lattice sites within $E_{\rm pa}$, since it
depends on the concentrations of the impurity complexes.  Hence the
Maxwell construction is exact only at zero temperature, and a full
statistical mechanical description is required at finite
temperatures. The standard plot of the defect formation energy against
the chemical potential found in the defect literature (see, for example,
Fig.\ 2 of Ref.\ \onlinecite{Zhang:PRL:1991}), suffers from the same
limitation.

\subsection{Application to N/O complexes in silicon}

\begin{figure}
\includegraphics*[width=8cm]{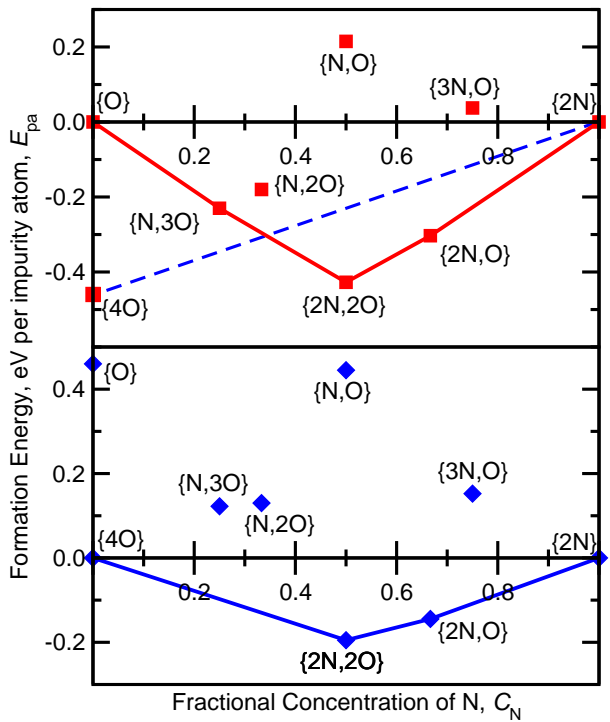}
\caption[]{(Color online) Maxwell construction for N/O complexes in
  silicon.  The upper panel shows the formation energy per impurity
  atom ($E_{\rm pa}$) for various complexes using the \{2N\} and \{O\}
  defects for the chemical potentials.  The stable complexes,
  \{N,3O\}, \{2N,2O\} and \{2N,O\} are found on the convex hull (solid
  red).  The blue (dashed) line is a tie-line between the values of
  $E_{\rm pa}$ for the \{4O\} and \{2N\} complexes.  If the O chemical
  potential is changed to that of the more stable \{4O\} complex, the
  stable complexes become those which both lie on the solid-red convex
  hull and also lie below the blue dashed line. This construction
  shows the stable complexes when the O chemical potential is changed
  to the energy of \{4O\}, since only the complexes with formation
  energies below the tie line (\{2N,2O\} and \{2N,O\}) are stable.
  This is confirmed by replotting the energies (blue diamonds) in the
  lower panel using \{4O\} as the chemical potential.  Only two
  structures lie on the new convex hull (solid blue line).}
\label{Fig:NO}
\end{figure}

We use N/O complexes in silicon to illustrate our approach and outline
the advantages of the method.  Fig.\ \ref{Fig:NO} shows a convex hull
containing N/O complexes from a previous study. \cite{morris:PRB:2009}
The hull shown in the upper panel of Fig.\ \ref{Fig:NO} is constructed
as follows.  We choose the chemical potentials for the N and O atoms
to be the same as in our previous study.\cite{morris:PRB:2009} The
chemical potential for N is obtained from the energy of the \{2N\}
defect as
\begin{equation}
  \mu_{\rm N}(\{2{\rm N}\}) = \frac{1}{2} \left( E_{\mathrm{D}}(\{2{\rm N}\}) 
    - E_{\mathrm{H}} \right),
\label{eqn:mu_{2N}}
\end{equation}
while for O it is obtained from the energy of the \{O\} defect as 
\begin{equation}
  \mu_{\rm O}(\{{\rm O}\}) = E_{\mathrm{D}}(\{{\rm O}\}) - E_{\mathrm{H}}.
\label{eqn:mu_{O}}
\end{equation}
These energies are plotted on the diagram at the points ($E_{\rm
  pa}$=0,$C_{\rm N}$=0) and (0,1).  The solid-red convex hull is then
drawn as a series of straight-line segments between (0,0) and (0,1)
passing through the squares representing the energies per impurity
atom $E_{\rm pa}$ in such a way that none of the defects lie below the
hull.  Only the complexes which lie on the hull are stable.
The \{O\}, \{N,3O\}, \{N,2O\}, \{2N,2O\} and \{2N,O\} defects are all
stable since they lie on the tie line of the convex hull, but the
\{N,2O\} complex is not stable and the energy would be lowered if
these defects disproportionated such that

\begin{equation}
  4\{{\rm N},2{\rm O}\} \rightarrow 2\{{\rm N},3{\rm O}\} 
  + \{2{\rm N},2{\rm O}\}.
\end{equation}

The dashed blue line in the upper panel of Fig.\ \ref{Fig:NO}
indicates how the situation would be modified if the chemical
potential for O is changed to the energy of the much more stable
\{4O\} complex.  The modified complex hull is redrawn as the
solid-blue line in the lower panel of Fig.\ \ref{Fig:NO}.  It is clear
from either the upper or lower panels that the \{N,3O\} defect is no
longer stable under these circumstances and the energy would be
lowered if they disproportionated such that
\begin{equation}
  2\{{\rm N},3{\rm O}\} \rightarrow \{4{\rm O}\} + \{2{\rm N},2{\rm O}\}.
\end{equation}
These results are entirely consistent with the analysis reported in
Ref.\ \onlinecite{morris:PRB:2009}.  The graphical representation is
useful because it shows which defects are stable for a particular
choice of chemical potentials, and it also allows an easy
visualization of the effects of altering the chemical potentials.  The
convex hull also shows the amount by which the energy is lowered by
the disproportionation of defect complexes.

\subsection{Native defects}

Our approach is easily extended to account for a crystal containing
$n_v$ native vacancies $V$ which form complexes
\{$n_{\alpha}\alpha$,$n_{\beta}\beta$,$n_vV$\} or $n_I$ interstitials
$I$ forming complexes \{$n_{\alpha}\alpha$,$n_{\beta}\beta$,$n_II$\}.
The native defects may be included within this framework, not as new
``impurities species'', but as a change in the definition of the host
material.  The chemical potential of an atom of the host material
added at a surface of the crystal is equal to the cohesive energy of
an atom in the perfect bulk crystal since and, as long as the surface
is large, the surface energy is unchanged by adding an atom at a
``kink site'' on a surface terrace.  Hence in the case where the
vacancy is absent prior to the addition of impurities, $\mu_{\beta}$
is the chemical potential of each host species ${\beta}$ in the
perfect crystal.  If the vacancy is present prior to the addition of
impurities, $\mu_{\beta}$ is the chemical potential of host species
${\beta}$ in a crystal containing the vacancy.  Note that it is not
possible to describe more than one type of vacancy in this approach
unless each type is treated as a separate species.  The same applies
for interstitials.

\section{Study of H/Li complexes in silicon}
\label{Sect:example}

\subsection{H and Li in silicon}

Hydrogen is a common impurity in silicon which binds to a variety of
defect complexes, including those containing oxygen and
nitrogen.\cite{morris:PRB:2009} It is difficult to determine the
concentration of hydrogen in silicon samples although it is likely to
be present at virtually every stage of manufacture and it may be
incorporated at levels of up to about
$10^{20}$\,cm$^{-1}$.\cite{amv:1992} The role of hydrogen impurities
in semiconductors has been reviewed by
Estreicher.\cite{estreicher:MSE:1995} In this article we present
results for H/Li impurities in bulk silicon.  Both
amorphous\cite{chevrier:JES:2009,chevrier:JES:2010} and
crystalline\cite{chevrier:JAC:2010} phases of the Li-Si system have
been studied using density-functional-theory (DFT) methods, giving
voltages which agree with experiment to within 0.1\,V.  Wan \emph{et
  al.}\cite{wan:JPC:2010} studied lithium defects in silicon using DFT
and found the most stable position for a single lithium impurity to be
at the tetrahedral ($T_d$) site, which is preferred to substitution on
a silicon site.  Kim \etal\cite{kim:JPCC:2010} performed a detailed
study of the electronic structure of a lithium atom in bulk silicon,
concluding that the presence of lithium weakens the nearby Si--Si
bonds.

\subsection{DFT calculations}

We have predicted structures of lithium impurities in silicon using
DFT and the \emph{ab initio} random structure searching method
(AIRSS).\cite{pickard:PRL:2006:silane,pickard:JPC:airss_review} In
this method randomly generated structures are relaxed to a minimum in
the energy.  This approach has been successful in predicting the
structures of point defects in
semiconductors\cite{morris:PRB:2008,morris:PRB:2009} and
ceramics,\cite{mulroue:PRB:2011} as well as high-pressure phases of
materials such as solid silane,\cite{pickard:PRL:2006:silane}
hydrogen,\cite{pickard:NP:2007:hydrogen} and
lithium.\cite{pickard:PRL:2009:lithium} Each AIRSS run was performed
at a fixed stoichiometry.  Searches were performed at several
stoichiometries and the results were analyzed using the convex hull
construction.

We used simulation cells containing 1, 2, 3 or 4 lithium atoms, 1, 2,
3 or 4 lithium atoms plus 1 hydrogen, atom and 1, 2 or 3 lithium plus
2 hydrogen atoms.  The initial structures were chosen in a similar
fashion to those in our earlier study of H/N/O
defects.\cite{morris:PRB:2009} A small hole was made in the host
crystal by removing a silicon atom from the unit cell and placing the
appropriate number of lithium atoms and a silicon atom randomly within
the hole.  The configurations were then relaxed using the DFT forces.
Some of the starting configurations relaxed to structure in which two
or more separate defects were present, and these structures were
discarded.

The plane-wave basis-set DFT code \textsc{castep}\cite{CASTEP:ZK:2004}
was used to calculate ground-state total energies and optimize the
geometries of the candidate structures.  We used ``on-the-fly''
ultrasoft pseudopotentials\cite{Vanderbilt:PRB:1990} and the
PBE\cite{perdew:PRL:1996} parameterization of the generalized gradient
approximation to the exchange-correlation functional.  The initial
searches were performed with simulation cells containing 32 silicon
atoms plus impurity atoms, sampling the Brillouin zone using the
$2\times2\times2$ multi-\emph{k}-point
generalization\cite{Rajagopal:PRL:1994,Rajagopal:PRB:1995,morris:PRB:2008}
of the Baldereschi scheme.\cite{Baldereschi:PRB:1973} The basis set
contained all plane-waves with energies up to 300\,eV.

The lowest-energy structures from each search were then further
relaxed at a higher level of accuracy.  We used simulation cells
containing 256 silicon atoms and a basis set with plane-waves up to
500\,eV, a harder Li pseudopotential and a standard $2\times2\times2$
Monkhorst-Pack grid of \emph{k}-points.  The formation energies were
then calculated using these structures but with a larger
$4\times4\times4$ Monkhorst-Pack grid of \emph{k}-points.  We also
performed some calculations in which we allowed spin polarization to
occur, but no significant effects were found.

\subsection{Results for H/Li in silicon}

\begin{table}[tbp]
{\centering \begin{tabular}{lr}
\hline
\hline 
Defect & $E_f$ (eV) \\
\hline  
\{3Li\} &   0.422 \\
\hline
\{H,3Li\}$^*$ & 0.103 \\
\{H,3Li\} &  -0.139 \\ 
\hline
\{H,2Li\}$^{**}$ &  0.321   \\
\{H,2Li\}$^{*}$ & 0.301 \\
\{H,2Li\} &   0.101  \\
\hline
\{2H,3Li\}$^{**}$ &   0.114 \\
\{2H,3Li\}$^{*}$ & -0.010  \\
\{2H,3Li\} & -0.239   \\
\hline
\{H,Li\}$^{**}$  &   0.474 \\
\{H,Li\}$^{*}$  & 0.472 \\ 
\{H,Li\} &  0.401 \\
\hline
\{2H,2Li\} &  -0.028  \\
\hline
\{2H,Li\}$^{*}$ &0.413\\
\{2H,Li\} &  -0.076 \\
\hline \hline
\end{tabular}\par}
\caption[Title]{
  Formation energies, $E_f$, of various H/Li complexes in silicon.
  The chemical potential for H is obtained from the energy of an 
  H$_2$ molecule at the $T_d$ site of bulk silicon and the chemical 
  potential for Li is obtained from the energy of a Li atom at the 
  $T_d$ site.  Single asterisks (*) and double asterisks (**) denote 
  the first and second metastable defects, respectively, for a given 
  defect stoichiometry.  The same information is represented graphically 
  in Fig.\,\ref{Fig:just_the_hull}. } 
\label{Tab:HLi_complexes}
\end{table}

\begin{figure}
\includegraphics*[width=80mm]{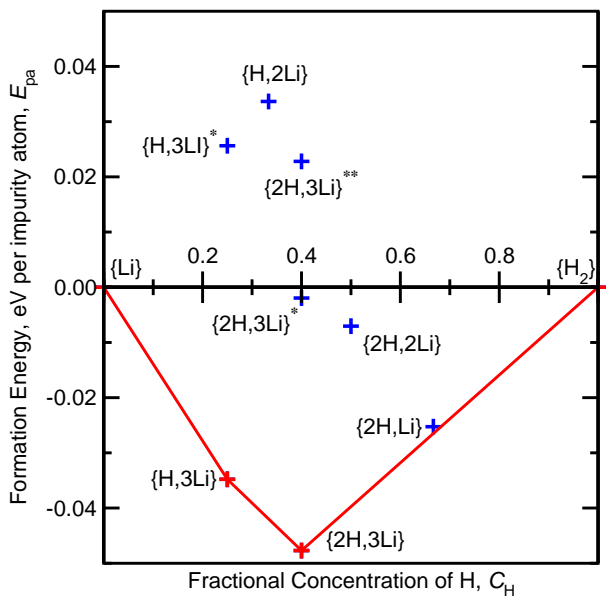}
\caption[]{(Color online) Convex hull for Li/H complexes in silicon.
  Formation energy per impurity atom plotted against the fractional
  concentration $C_{\rm H}$ of hydrogen.  The plus signs ($+$)
  represent defect complexes.  The chemical potentials for Li (\{Li\})
  and H (\{H$_2$\}) are plotted at (0,0) and (0,1), respectively.
  Metastable complexes of a given stoichiometry are denoted by
  asterisks ($^*$).  The formation energies of defect complexes are
  indicated by blue (dark grey) crosses and the predicted stable
  defects by red (mid-grey) crosses.  The red (mid-grey) lines show
  the formation energy per impurity atom of the (mixtures of) stable
  defects present as a function of $C_{\rm H}$.}
\label{Fig:just_the_hull}
\end{figure}

\begin{figure}
\includegraphics*[width=4cm]{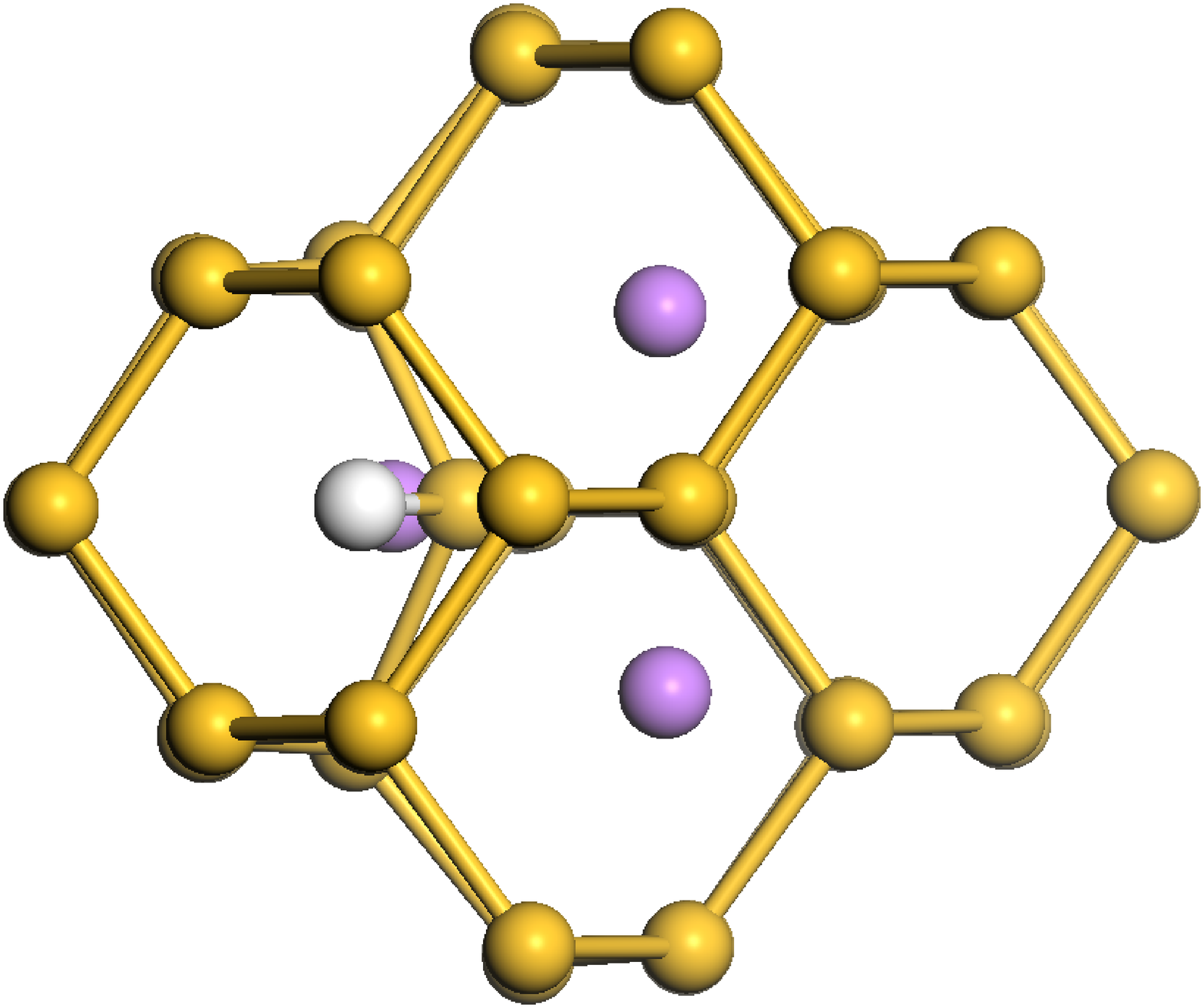}
\includegraphics*[width=4cm]{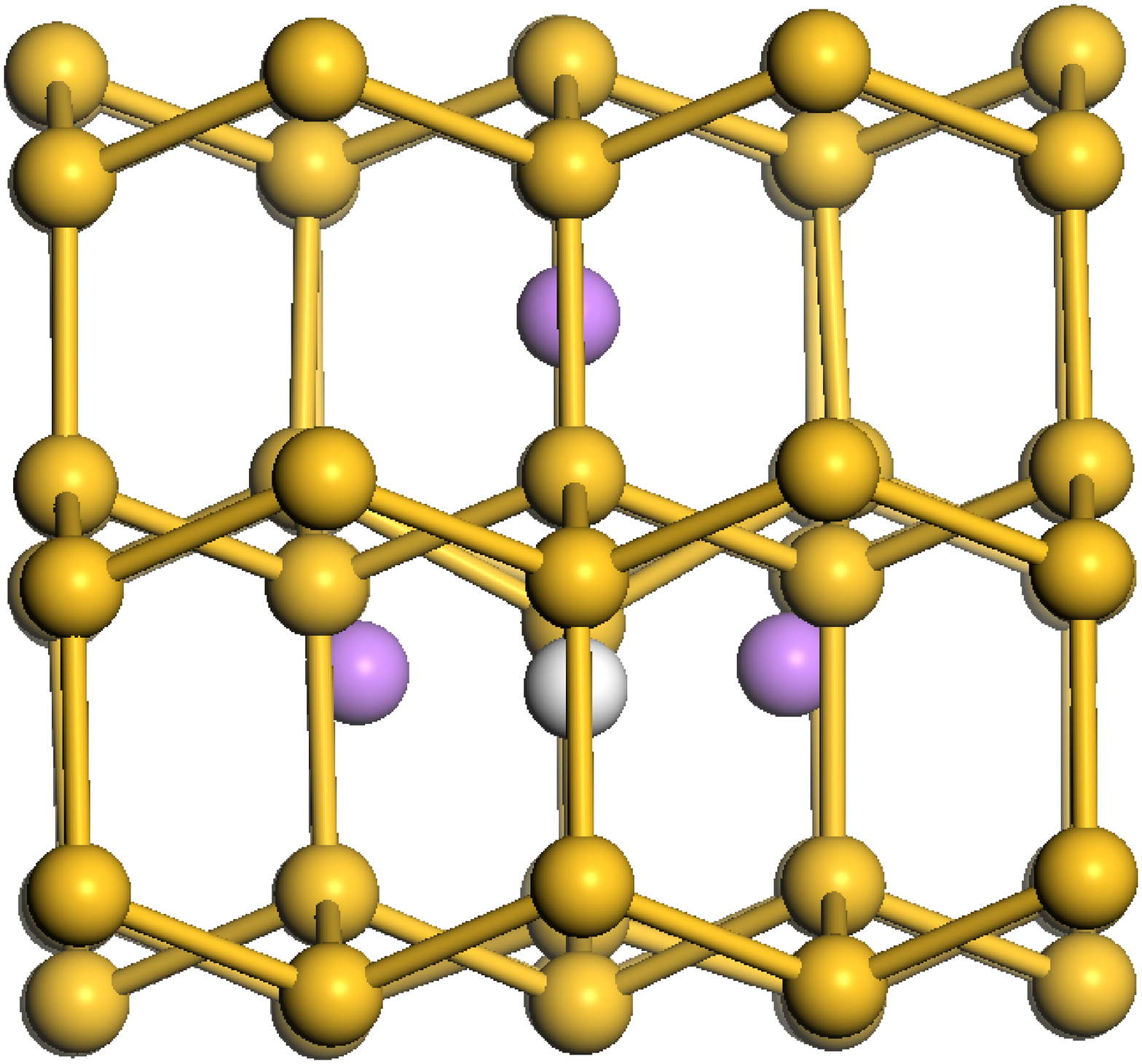}
\caption[]{(Color online) Two views of the \{H,3Li\} defect in bulk
  silicon.  Silicon atoms are shown in yellow (light grey), lithium in
  purple (dark grey) and hydrogen in white.  This defect is predicted
  to be stable for $0<C_{\rm H}<0.4$.  The lithium atoms form a
  three-membered ring in a void created by breaking a Si--Si bond and
  a hydrogen atom bonds to the displaced silicon atom.}
\label{Fig:3LiH}
\end{figure}
\begin{figure}
\includegraphics*[width=4cm]{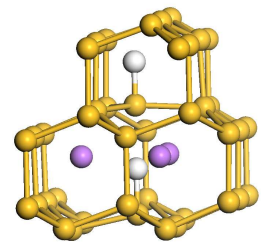}
\caption[]{(Color online) The \{2H,3Li\} complex in bulk silicon.
  Silicon atoms are shown in yellow (light grey), lithium in purple
  (dark grey) and hydrogen in white.  This defect is predicted to be
  stable for $0.25<C_{\rm H}<1.0$. It is similar to the \{H,3Li\}
  defect shown in Fig.\,\ref{Fig:3LiH}, except that the additional
  hydrogen atom terminates the dangling bond on the displaced silicon
  atoms.  The lithium atoms form a three-membered ring in the void
  created by breaking a Si--Si bond.  A hydrogen atom sits between
  these two silicon atoms. }
\label{Fig:3Li2H}
\end{figure}

We model the initial stages of the incorporation of Li atoms into bulk
silicon as the formation of defect complexes with varying
concentrations of H and Li atoms. We assume that the Li atoms are
supplied from bulk body-centered-cubic Li metal.  The \{Li\} defect in
silicon has a formation energy of 0.21 eV relative to bulk Li in
accordance with the observed overpotential during Li insertion
reactions in bulk silicon. \cite{obrovac:JES:2007} We assume that the
Li atoms initially form \{Li\} defects in the bulk silicon and we
investigate whether chemical reactions between Li and H atoms in bulk
silicon are energetically favorable.  We choose the \{Li\} defect to
define our chemical potential because it is the lowest energy defect
that we have found in silicon and because a lithium defect of the same
($T_d$) symmetry has been observed in electron paramagnetic resonance
experiments in bulk silicon.\cite{watkins:PRB:1970} A metastable
\{3Li\} complex with $C_{3v}$ symmetry was also found with formation
energy relative to \{Li\} of 0.14\,eV per atom, but this is too high
in energy to appear on Fig.\ \ref{Fig:just_the_hull}.  The hydrogen
chemical potential was obtained as in previous
studies\cite{morris:PRB:2008,morris:PRB:2009} from a H$_2$ molecule at
the $T_d$ site in silicon.  A more complete model would require the
inclusion of other defects, such as vacancies and vacancy complexes.

Single-atom Li defects are favored when only Li impurities are
present, but the convex hull plotted in Fig.\,\ref{Fig:just_the_hull}
shows that the addition of hydrogen causes the lithium atoms to
cluster.
For $0<C_{\rm H}<0.25$ the \{H,3Li\} defect forms and any
additional lithium atoms occupy $T_d$ sites.  The structure of the
\{H,3Li\} defect is pictured in Fig.\,\ref{Fig:3LiH}.  Both the
\{H,3Li\} and \{2H,3Li\} complexes are present in the range
$0.25<C_{\rm H}<0.40$.  The \{2H,3Li\} defect is present when $C_{\rm
  H}>0.40$, and any excess hydrogen forms H$_2$ molecules at the $T_d$
site.  The \{H,3Li\}, \{2H,3Li\} and \{2H,Li\} defects resemble a
Chadi H$^{*}_2$ defect\cite{chadi:APL:2003} with the lithium atoms
close by.  The metastable H$^{*}_2$ complex found by
Chadi\cite{chadi:APL:2003} has a formation energy relative to a H$_2$
molecule at the $T_d$ site of 0.1\,eV per H
atom.\cite{morris:PRB:2008} The \{2H,Li\} defect could be present when
thermal effects are taken into account, due to its close proximity to
the tie line at $C_{\rm H}=0.66$.

\section{Conclusions}
\label{Sect:conclusions}

We have used convex hull diagrams based on the Maxwell construction to
visualize the energetics of defect complexes.  The graphical
representation shows the stable defects for a particular choice of
chemical potentials and the relative stabilities of the defect
complexes.  It allows visualization of the changes in the stability of
the defects when the chemical potentials are changed.  The approach
was illustrated using data for N/O complexes in silicon from a
previous study.\cite{morris:PRB:2009} The method can easily be
extended to crystals containing charged species, native vacancies or
interstitials, which is an area we are currently investigating. 

We also studied H/Li complexes in silicon, which may be relevant to
understanding how the first few Li atoms enter crystalline silicon.
This extended the previous analysis of lithium in silicon anodes to
include its interaction with the hydrogen that is introduced into the
silicon during manufacture.  When only lithium impurity atoms are
present the most stable defect is \{Li\}, but the presence of hydrogen
atoms causes the lithium atoms to form H/Li clusters containing three
lithium atoms.  We found that the \{H,3Li\} and \{2H,3Li\} complexes
are stable and that the \{2H,Li\} complex is nearly stable.  Hydrogen
present in the bulk favors bonding covalently to self-interstitial
defects,\cite{morris:PRB:2008} but once the silicon defects are
saturated, hydrogen will bind to lithium complexes, further
stabilizing them.  The combination of AIRSS and the Maxwell
construction and convex hull diagrams is a powerful tool for analyzing
the energetics of point defects in materials.

\begin{acknowledgments}
  We are grateful to Marshall Stoneham, Tony Harker, Dave Bowler and
  Mike Gillan for fruitful discussions.  This work was supported by
  the Engineering and Physical Sciences Research Council (EPSRC) of
  the U.K.\ Computational resources were provided by the University
  College London Research Computing service.
\end{acknowledgments} 

\bibliographystyle{h-physrev}

\end{document}